\documentclass[aps,pra,twocolumn,amsmath,amssymb,floats,floatfix,showpacs]{revtex4}

\usepackage{graphicx}% Include figure files
\usepackage{dcolumn}% Align table columns on decimal point

\begin{document}

%\draft

\title{Driven harmonic oscillator as a quantum simulator for open systems}

\author{Jyrki Piilo}

\affiliation{
Department of Physics, University of Turku, 
FI-20014 Turun yliopisto, Finland}

\author{Sabrina Maniscalco}
\affiliation{
Department of Physics, University of Turku, 
FI-20014 Turun yliopisto, Finland}

\date{\today}

\begin{abstract}
We show theoretically how a driven harmonic oscillator can be used as a quantum simulator
for the non-Markovian damped harmonic oscillator. In the general
framework, our results demonstrate the possibility to use a closed system as
a simulator  for open quantum systems.
The quantum simulator is based on sets of controlled drives
of the closed harmonic oscillator with appropriately tailored electric
field pulses. The non-Markovian dynamics of the damped harmonic
oscillator is obtained by using the information about the spectral density
of the open system when
averaging over the drives of the closed oscillator. 
We consider single trapped ions as a specific physical implementation 
of the simulator, and we show how the simulator approach reveals
new physical insight into the open system dynamics, 
e.g.~the 
characteristic quantum mechanical non-Markovian oscillatory behavior
of the energy of the damped oscillator, usually obtained
by the non-Lindblad-type master equation, can have
a simple semiclassical interpretation.
\end{abstract}


\maketitle

\section{Introduction}

Recent years have witnessed a considerable experimental
and theoretical progress in quantum information 
science~\cite{Nielsen,Stenholm,EuDocu}. A major factor for the rapid development
has been the increasing ability to control and engineer
the quantum states of single and few particle systems.
The development
of the field is expected to continue in fast pace and some applications,
e.g.~for quantum cryptography, are already commercially
available. One of the most active contemporary subfields of 
quantum information science
and of nanotechnology is the study of  quantum 
simulators~\cite{Sorensen,Wineland1,Zoller1, Milburn, Stoof,EuDocu}.

Simulating quantum systems by classical computers
remains a challenging task due to the inherent inability of classical systems
to incorporate the quantum features in an efficient way.
The development of quantum simulators, i.e.~a controllable quantum system 
imitating the behavior of other quantum systems of interest,
holds the promise for a generation of powerful means and devices
to study quantum systems and provides additional ways
to gain new insight into their peculiar quantum features. 

A notable aspect of recent proposals for quantum simulators
is their ability to cross the usual boundaries
between different fields of physics~\cite{Zoller1, Milburn, Stoof}, 
e.g.~a certain configuration
of cold atoms trapped in optical lattices has been recently proposed to simulate
superstrings~\cite{Stoof}. These, as well as more
\lq\lq conservative\rq\rq~quantum
simulators~\cite{Sorensen,Wineland1}, 
are often based on cold trapped atoms and gases because of 
the recently developed abilities to manipulate their properties
in an extremely precise way. 

In this paper we develop a quantum simulator scheme
for open quantum systems exploiting the recent progress
in trapped ion technologies \cite{Wineland2}. The motivation to develop quantum simulators for
open systems stems especially from their central role in the progress
towards fault-tolerant quantum information processing. Indeed, 
decoherence, which is a key issue for successful quantum
gate implementation, is a topic typically studied in terms
of open quantum systems \cite{Breuer}. 

We focus on a quantum simulator scheme for a 
damped harmonic oscillator which is a
paradigmatic model of open quantum systems~\cite{Breuer}
and has
a wide variety of applications ranging, e.g., 
from quantum optics~\cite{Qo} and nuclear physics~\cite{Np} to 
chemistry~\cite{Ch}.
We show how a driven harmonic oscillator
can be used to mimic the behavior of the damped harmonic
oscillator. In other words, we demonstrate theoretically the possibility
to use a closed system as a quantum simulator for open
quantum system. 

Our quantum simulator is based on the controlled driving of a closed
harmonic oscillator by sets of appropriately tailored electric fields.
We show how the sets of drives in different
frequency regimes give rise to various types of dynamical
features of the driven oscillator, and how these
shed light on the effects that the various
parts of the environment spectrum have
on the open system dynamics to be simulated.
The total non-Markovian dynamics of the damped harmonic
oscillator is obtained by using the information
about the spectral density 
of the open system when
averaging over the sets of drives of the closed oscillator.

A well known example of experimental realization of
quantum harmonic oscillator is a single trapped ion \cite{Wineland2}
and we discuss the implementation of the simulator with this
system. The effects of random electromagnetic fields on
the trapped ion dynamics has been recently 
studied~\cite{Lamoreaux,James,Budini}.
In the usual
approaches, the system dynamics is  calculated averaging
over known stochastic properties of the environment
fields. Our approach 
is taken from the opposite direction.
We ask if it is possible to drive the closed oscillator in a controlled way
so that the effect mimics the open system dynamics.
Our results highlight the possibility to simulate artificial, engineered reservoirs,
by controlled sets of drives.
In general, reservoir engineering is emerging
as an active contemporary research topic for the 
control and the fundamental study of open quantum 
systems~\cite{Zoller2,ReservoirEng,WinelandPRA,Zoller3}.

The quantum simulator approach,
already at a theoretical level, makes it possible to gain new insight into the complex
non-Markovian dynamics of the damped oscillator. 
The simulator approach allows to identify 
the origin of the different dynamical features of the non-Markovian
dynamics of the open system and sheds light 
on the role that the phase of the random environmental fields
plays in the system dynamics. Furthermore we show how the implementation
of the quantum simulator with a trapped ion allows to interpret a typical 
quantum mechanical behavior, usually described with a non-Lindblad-type master equation, by semiclassical means. We also discuss the implications
of the quantum simulator approach towards the continuos measurement
scheme interpretation of the quantum trajectories
for non-Markovian open systems~\cite{MeasurementScheme}.

The paper is organized as follows. Section \ref{Sec:Drive}
introduces the model of a closed harmonic oscillator 
driven by controlled fields and sets the framework for the
quantum simulator approach  presented in 
Sec.~\ref{Sec:QuaSim}. Here, the damped oscillator model
to be simulated, is presented,
and we furthermore discuss the new insight 
that the simulator
approach brings to light.
Section~\ref{Sec:IonImp} shows how the quantum simulator can be implemented
with a trapped ion and the discussion concludes the paper
in Sec.~\ref{Sec:DisCon}.

\section{Driven oscillator dynamics}\label{Sec:Drive}

\subsection{The basic scheme: pulsed driving}

We begin by considering the dynamics of a closed quantum harmonic oscillator
which is driven by a single
frequency field for a set of runs which we label by the frequency $\omega$ of the 
field. In each run of the set $\omega$, the duration
of the applied field changes. In other words, a single set of runs
with frequency $\omega$ consists of drives with increasing duration
of the pulses mapping the oscillator dynamics as a function of time.
The frequency of the driving field changes
for different sets of runs and
depends on the type
of the environment and spectral density 
to be simulated.

It is important to emphasize that the dynamics of the closed oscillator is
not monitored (measured) in a continuos way.
Rather, the scheme for a single run of the set $\omega$ can be summarized
as follows:
(i) Switch on the driving field at time $t_1=0$. The state of the oscillator
may change suddenly due to a sudden switch on of the driving field. 
(ii) Driving of the oscillator with external field of frequency $\omega$ for $0<t<t_2$.
(iii) Switch off of the driving field at $t=t_2$.
The state of the oscillator may change suddenly due to a sudden switch off of the driving field.
With the set of pulses, with increasing $t_2$ for the members
of the set $\omega$, one maps the
oscillator dynamics. The key ingredient for the simulator,
for the system we study,
is related to the effects of switching on and off of the field.

The Hamiltonian of a 
quantum harmonic oscillator driven by a time dependent 
periodic force is~\cite{Gardiner96a}
\begin{equation}
H=\hbar\omega_0\left(a^{\dag} a+\frac{1}{2}\right)+
\hbar F(t) (a+a^{\dag}).
\label{Eq:H}
\end{equation}
Here, $\omega_0$ is the frequency of the oscillator, $a^{\dag}$ ($a$)
is the creation (annihilation) operator of the energy quanta of the oscillator,
and the periodic driving force is
\begin{equation}
F(t) = \frac{A\cos(\omega t + \varphi)}{\sqrt{2m\hbar\omega_0}},
\end{equation}
where $A$ describes the amplitude and $\omega$ the oscillation
frequency of the periodic force, $m$ is the mass of the oscillator,
and $\varphi$ is the phase of the driving field.

\subsection{Off-resonant driving}

We assume, for simplicity, that
the oscillator is initially in the ground state.
In this case it can be shown that the heating
function $\langle n \rangle = \langle a^{\dag} a\rangle$,
for $\omega\neq\omega_0$, can be written
\begin{eqnarray}
\langle n \rangle (t,\omega)&=& 
|\alpha|^2 
\left[\left(\frac{\omega}{\omega_0}\right)^2-1 \right]^{-2}
\left[1+\cos^2(\omega t) 
\nonumber \right.\\
&&
- 2 \cos(\omega t) \cos(\omega_0 t)
-2\frac{\omega}{\omega_0}\sin(\omega t)\sin(\omega_0 t)
\nonumber \\
&&
\left.
+\left(\frac{\omega}{\omega_0}\right)^2\sin^2(\omega t)
\right],
\label{Eq:Drive}
\end{eqnarray}
where the phase of the field is chosen
for convenience $\varphi = 0$
and $\alpha=A / \sqrt{2m\hbar\omega_0}$.

Figures \ref{Fig:Freqs} (a), (b), and (c) show examples of the time evolutions
of $\langle n\rangle$ for $\omega/\omega_0=0.0,~0.1,~0.2$.
In the following two subsections we discuss in detail the heating function
dynamics for constant field drive ($\omega=0$) and in the adiabatic regime
($\omega\ll\omega_0$). 
In these frequency ranges, the dynamical features of the driven
oscillator heating function allow to gain interesting insight into the non-Markovian
dynamics of the open system (to be discussed in Sec.~\ref{Sec:QuaSim}). 

\subsubsection{Constant field}

When driving the system
with a constant field $\omega=0$,
the heating function displays sinusoidal
oscillations, see Fig.~\ref{Fig:Freqs} (a).
This can be explained
in the following way: (i) Switching on the driving field at time $t_1=0$.
A suddenly switched on constant force displaces the ground state
of the oscillator creating a coherent state
$D(\alpha) |0\rangle = |\alpha\rangle$ (ii) Time evolution between $t_1 < t < t_2$.
The coherent state oscillates in the trap according to the free evolution
(no change in $\langle n \rangle$).
(iii) Switching off the field at $t=t_2$.
The initial displacement of (i) is reversed [$D(-\alpha)$]. 
Moreover, the effect of the second displacement depends on the phase of the oscillation
of the coherent state.
 
The final state of the system at time $t_2$ (after the second 
displacement) is
\begin{equation}
|\psi(t)\rangle = |-\alpha+\alpha \exp(-i \omega_0 t)\rangle,
\end{equation}
and the heating function becomes
 \begin{equation}
\langle n \rangle (t) = |-\alpha+\alpha \exp(-\imath  \omega_0 t)|^2 =
2|\alpha|^2 (1-\cos\omega_0 t).
\label{Eq:ConDr}
\end{equation}
The second displacement brings
the coherent state back to the ground state after full oscillation periods
whereas half the period of oscillation of the coherent state during $t_1<t<t_2$ causes
a total displacement of $2\alpha$.
The heating function consequently oscillates between $0$ and $4 |\alpha|^2$.
Setting $\omega=0$ in the general result of Eq.~(\ref{Eq:Drive}) matches the simple
result [Eq.~(\ref{Eq:ConDr})] and confirms this interpretation.

\subsubsection{Driving with small frequency in adiabatic regime}

Let us consider now the case $0<\omega\ll\omega_0$.
Since the driving frequency is much smaller than the oscillator
frequency, the oscillator follows adiabatically the changes
of the force.
Figures \ref{Fig:Freqs} (b) and (c) display examples of drivings in this regime for
$\omega/\omega_0=0.1,~0.2$.  

Also here, the switches on and off of the drive field are thought to happen
instantaneously compared to the oscillator dynamics.
The simple scheme of the previous subsection
is replaced with:
(i) Switching on the field at time $t_1=0$.
As before, the switched on field displaces the ground state
of the oscillator creating a coherent state
$D(\alpha) |0\rangle = |\alpha\rangle$
(ii) Time evolution between $t_1 < t < t_2$.
The coherent state oscillates in the trap according to the free evolution (no change in $\langle n \rangle$).
Moreover, the periodic driving force with frequency $\omega$
changes slowly. The oscillator follows adiabatically the consequent slow motion of the 
oscillator center since $\omega_c\ll\omega_0$.
(iii) Switching off the field at $t=t_2$.
Because of the slow oscillation of the driving force,
the magnitude of the second displacement depends now also on driving frequency $\omega$ 
and on time $t$: $D(-\cos(\omega t) \exp(-i\omega_0t)\alpha)$. 

Following this scheme, the state of the oscillator after the second displacement can
be written 
\begin{equation}
|\psi(t)\rangle = |e^{-i\omega_0 t}\alpha-\cos(\omega t)\alpha\rangle,
\label{Eq:Psi}
\end{equation}
and the heating function becomes
\begin{equation}
\langle n \rangle(t, \omega) = |\alpha|^2 \left[1+\cos^2(\omega t) - 2 \cos(\omega t) \cos(\omega_0 t)\right].
\label{Eq:Nw}
\end{equation}
This matches the general result of Eq.~(\ref{Eq:Drive}) when the terms
containing $\omega/\omega_0\ll1$ are neglected
(adiabatic limit) and confirms the interpretation presented here.

\subsection{On resonant driving}

\begin{figure}[tb]
\centering
\includegraphics[scale=0.4]{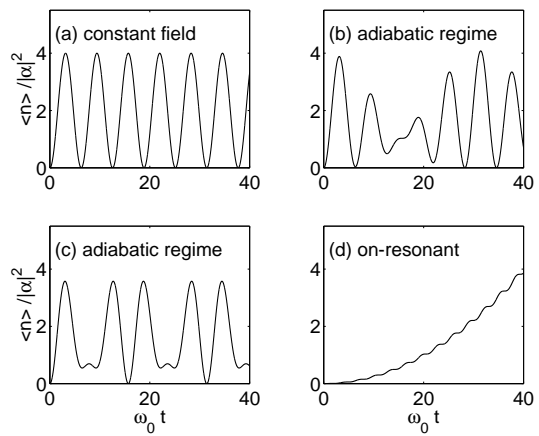}
\caption{\label{Fig:Freqs} 
Time evolution of the heating function
of the closed driven oscillator in three different frequency regimes for four different driving frequencies: (a) $\omega/\omega_0=0$,  
(b) $\omega/\omega_0=0.1$,  (c) $ \omega/\omega_0=0.2$,
(d) $\omega/\omega_0=1.0$. 
Please note that we have multiplied the result
in (d) by a factor of $0.01$ (a weight factor given
by the spectral density we use in Sec.~\ref{Sec:QuaSim}).
}
\end{figure}

The equation (\ref{Eq:Drive}) for the heating function is not well defined 
for $\omega=\omega_0$. Beginning from the Hamiltonian
of Eq.~(\ref{Eq:H}) one can show that on-resonant driving
of the oscillator gives for the heating function (for convenience the 
phase of the field is set $\varphi=0$)
\begin{equation}
\langle n\rangle(t, \omega_0) = \frac{1}{4}|\alpha|^2 
\left[\omega_0^2t^2 + \omega_0t \sin(2\omega_0t)
+\sin^2\left(\omega_0t\right)\right].
\label{Eq:Nw0}
\end{equation}
Figure \ref{Fig:Freqs} (d) displays
an example of the heating function dynamics 
in this case. We note
that the oscillatory terms play a minor role
contrary to the off-resonant constant field and adiabatic cases.

%%%%%%% FIG %%%%%%%%%
%%%%%%%% FIG %%%%%%%%%

\section{Driven oscillator as a quantum simulator}\label{Sec:QuaSim}

In the previous section we have shown the main
dynamical features of the pulsed driven
oscillator in various frequency ranges of the driving force
(constant force, adiabatic regime, on-resonant driving). 
We show now that by combining the dominant
dynamical features from the three main frequency regimes,
one can construct a quantum simulator for an open quantum
system. We begin by recalling the basic features of the system
to be simulated.

\subsection{System to be simulated: damped harmonic oscillator}

The damped harmonic oscillator is one of the paradigmatic models 
used to describe the dynamics of open quantum systems~\cite{Breuer}.
We briefly recall the form of the master equation for the reduced system
density matrix which allows to describe the non-Markovian dynamics of the damped oscillator.
For a more detailed presentation, see e.g.~Ref.~\cite{Maniscalco04a}
and references therein.

The dynamics of a harmonic oscillator linearly coupled to a
quantized reservoir, modelled as an infinite chain of quantum
harmonic oscillators, is described, in the secular approximation, by
means of the following generalized master equation
\cite{Intravaia03a,Maniscalco04a}
\begin{eqnarray}
&&\frac{ d \rho(t)}{d t}= \frac{\Delta(t) \!+\! \gamma (t)}{2}
\left[2 a \rho(t) a^{\dag}- a^{\dag} a \rho(t)  - \rho(t)
a^{\dag} a \right]
\nonumber \\
&& +\frac{\Delta(t) \!-\! \gamma (t)}{2} \left[2 a^{\dag} \rho(t)
a - a a^{\dag} \rho(t) - \rho(t) a a^{\dag}
 \right], \nonumber \\
 \label{Eq:MQbm}
\end{eqnarray}
with $\rho(t)$ the reduced density matrix of the
system harmonic oscillator. The time dependent coefficients
$\Delta(t)$ and $\gamma(t)$ appearing in the master equation are
known as diffusion and dissipation coefficients, respectively
\cite{Intravaia03a,Maniscalco04a}. 

The diffusion coefficient
for high temperature reservoir  $k_{\rm B}T/\omega_0\gg1$ and to
second order in the dimensionless coupling constant $g$, can be written
\cite{Caldeira83a,Maniscalco04a}
\begin{eqnarray}
\Delta(t) &=& 2 g^2 k_{\rm B} T \frac{r^2}{1+r^2} \left\{ 1
- e^{-\omega_c t} \left[ \cos (\omega_0 t)\right.\right.
 \nonumber \\
&&
-  (1/r)  \left. \sin (\omega_0 t )\right] \big\}.
\label{eq:deltaHT}
\end{eqnarray}
Above, $r=\omega_c/\omega_0$ is the ratio between the environment
cut-off frequency $\omega_c$ and the oscillator frequency
$\omega_0$,  $k_{\rm B}$ is
the Boltzmann constant, and $T$ is the temperature. 
For the high temperature case 
the dissipation coefficient $\gamma\ll\Delta$ and therefore is neglected.

\begin{figure}[tb]
\centering
\includegraphics[scale=0.4]{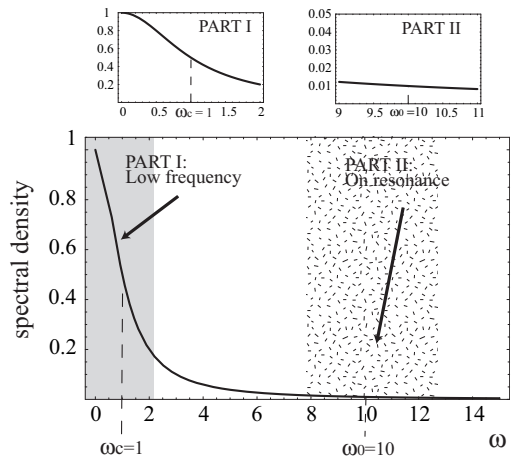}
\caption{\label{Fig:Spectrum1} 
Schematic view of the spectral density and the splitting of the spectrum
($r=0.1$).
Part I is the low frequency part which produces the sudden displacements
of the oscillator state
when switching on and off.
Part II is the on-resonant part which produces Markovian type heating.
}
\end{figure}

A typical environment of open systems is described by an Ohmic
reservoir spectral density with Lorentz-Drude cut-off~\cite{Weiss}
\begin{equation}
J(\omega)= \frac{2  \omega}{\pi} \
\frac{\omega_c^2}{\omega_c^2+\omega^2}, \label{Eq:SpecDen}
\end{equation}
for a schematic view, see Fig.~\ref{Fig:Spectrum1}.
The spectral distribution is given by
\begin{eqnarray}
I(\omega) &=& J(\omega) [n_e(\omega)+1/2] \nonumber \\
&=& \frac{ \omega}{\pi} \frac{\omega_c^2}{\omega_c^2+\omega^2}
\coth{(\omega/K T)} , 
\label{Eq:I}
\end{eqnarray}
where $n_e(\omega)$ is the population of the environment mode of frequency
$\omega$, and Eq.~(\ref{Eq:SpecDen}) has been used. For high $T$,
Eq.~(\ref{Eq:I}) becomes
\begin{equation}
I(\omega) = \frac{2 k_{\rm B} T}{\pi}
\frac{\omega_c^2}{\omega_c^2+\omega^2}.
\label{Eq:IHt}
\end{equation}

The central parameter $r=\omega_c/\omega_0$ describes how on-resonant
the oscillator is with the reservoir.  
When $r>1$, intensive part of the environment spectrum overlaps
with the oscillator frequency and the decay coefficient
$\Delta(t)>0$
for all times. Consequently, the master equation is of Lindblad-type.   
When $r<1$ (Fig.~\ref{Fig:Spectrum1}), the most intense part of the environment spectrum lies in small
frequency range and the on-resonant intensity  is small. 
In this case the decay coefficient $\Delta(t)$
acquire temporarily negative values
and the master equation is of non-Lindblad-type \cite{Maniscalco04a}.

%%%%%%%%%%%% FIG
%%%%%%%%%%%% FIG

The full solution of the the master equation
(\ref{Eq:MQbm}) can be found, e.g.,~in Refs.~\cite{Maniscalco04a,Maniscalco04c}.
For an initial Fock state $|n=0\rangle$ it turns out that, in the secular approximation, the state of the system at each
time $t$ is a thermal state with \lq\lq effective\rq\rq~temperature changing in time according to the behavior of $\langle n (t) \rangle$ (see Appendix A):
\begin{equation}
\rho_{\rm k,k} = \left[\frac{\langle n (t)\rangle}{\langle n (t)\rangle+1}\right]^k
\frac{1}{\langle n (t)\rangle+1}.
\label{Eq:RhokkAna}
\end{equation}
It is worth emphasizing that the short time non-Markovian effects are described by
the non-monotonic changes in $\langle n \rangle$ [see the exact result for
$\langle n \rangle$ in Fig.~3 (d)]. The non-Markovian regime
is then followed by Markovian linear heating and finally for long times
the system reaches a steady state in thermal equilibrium with its 
environment (see Ref.~\cite{OscNote} for a discussion of the origin of the oscillations
of the heating function in the non-Markovian regime and Ref.~\cite{Intravaia03b} for a discussion of the Markovian and thermalization regimes).

It is also possible to see from the master equation (9) that for an initial Fock
state the coherences vanish for all times
in the Fock-state basis.
This is due to the fact
that the equations of motions for the diagonal
elements $\rho_{k,k}(t)$
and off-diagonal elements $\rho_{k,k'}(t)$, with $k\neq k'$, decouple.
In particular this means that the equations of motion
for the off-diagonal elements $\rho_{k,k'}(t)$,  
do not depend on the diagonal elements $\rho_{k,k}(t)$.
Since for an initial
Fock state $\rho_{k,k'}(0)=\langle k | \rho (0) | k' \rangle=0$, 
the decoupling then implies that $\rho_{k,k'}(t)=0$
for all $t$ and $k\neq k'$. 

For an initial ground state in the weak coupling ($\langle n (t)\rangle \ll 1$) 
and for times much shorter than the thermalization time one obtains
\begin{eqnarray}
\rho_{\rm 0,0} &=&  \frac{1}{\langle n (t)\rangle + 1} \simeq 1 - \langle n (t)\rangle 
\nonumber
\\
\rho_{\rm 1,1} &\simeq& \langle n (t)\rangle.
\label{Eq:AnaRho}
\end{eqnarray}
The analytic solution for the heating function
and for $\langle n(0)\rangle=0$ is given by \cite{Maniscalco04a}
\begin{equation}
\langle n (t) \rangle = 
\int_0^t \Delta(t')  dt',
\label{Eq:Exact}
\end{equation}
where  $\Delta(t)$ has been given
in Eq.~(\ref{eq:deltaHT}).
The exact heating function dynamics, given by Eq.~(\ref{Eq:Exact}),
will be used as a benchmark for the theoretical quantum simulator
results in the following subsection.

\subsection{Simulating damped harmonic oscillator with driven oscillator}

We have calculated in Sec.~\ref{Sec:Drive} the heating function dynamics for
a driven oscillator. The oscillator is driven with single frequency
fields in a pulsed way, the frequencies covering the relevant  part 
of  the environment spectrum of the damped oscillator. We now
proceed to show how the complete damped harmonic oscillator dynamics is recovered
when the information about the spectral density
is used to average the driven oscillator results.
We first demonstrate that the simulator gives correct heating function
dynamics and then present the complete solution showing
that the density matrices of the simulator and of the damped harmonic oscillator match.

%%%%%%%%%%%% FIG
\begin{figure*}[!htb]
\centering
\includegraphics[scale=0.6]{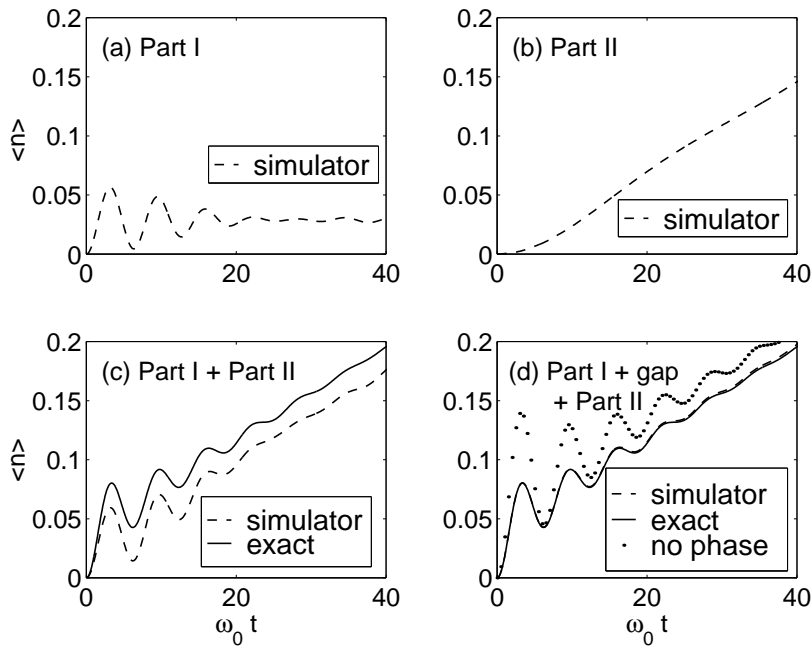}
\caption{\label{Fig:Parts} 
Comparison between the quantum simulator and the exact results
for the non-Markovian oscillatory dynamics.
The results demonstrate how the different parts
of the environment spectrum affect the dynamics.
(a) $0 \leq \omega/\omega_0 \leq 0.2 $, (b) $0.8 \leq \omega/\omega_0 \leq 1.2$,
(c) $0.0 \leq \omega/\omega_0 \leq 0.2$ and  $0.8 \leq \omega/\omega_0 \leq 1.2 $, 
(d) $0.0 \leq \omega/\omega_0 \leq 1.2$.
Moreover, the results in (d) show how the the random phase 
of the driving field influences the amplitude
of the oscillations of the heating function.
Due to the excellent match between the simulator
and the exact results in (d), it is difficult to see any difference
between the two.
}
\end{figure*}
%%%%%%%%%%%% FIG

\subsubsection{Heating function}

Equation (\ref{Eq:Drive}) 
describes the heating function
for off-resonant driving with $\varphi=0$.
When the phase of the driving field is taken into account,
the heating function can be written
\begin{widetext}
\begin{eqnarray}
\langle n \rangle (t, \omega, \varphi)
&=&
|\alpha|^2
\left[\left(\frac{\omega}{\omega_0}\right)^2-1 \right]^{-2}
\Big\{
\cos^2(\varphi)+\cos^2(\omega t+\varphi)
- 2  \cos(\varphi)\cos(\omega t + \varphi)\cos(\omega_0 t)  
-2\frac{\omega}{\omega_0}\sin(\omega t)\sin(\omega_0 t) \nonumber \\
&-&
2\left(\frac{\omega}{\omega_0}\right)^2\sin(\varphi)\sin(\omega t + \varphi)
\cos(\omega_0 t) 
+\left(\frac{\omega}{\omega_0}\right)^2
\left[\sin^2(\varphi)+\sin^2(\omega t + \varphi)\right]
\Big\}.
\label{Eq:WithPhi}
\end{eqnarray}
\end{widetext}
To obtain the final result, we average Eq.~(\ref{Eq:WithPhi}) over
the phase of the driving field and over the various
driving frequencies by using the spectral density as a weight
factor
\begin{eqnarray}
\langle n \rangle (t) &=& \frac{1}{N}\int_0^\infty d\omega 
\frac{1}{2\pi}\int_0^{2\pi}d\varphi
I(\omega)\langle n \rangle (t, \omega, \varphi).
\label{Eq:GenAve}
\end{eqnarray}
Averaging over the phase gives,
\begin{eqnarray}
\langle n \rangle (t) &=& \frac{|\alpha|^2}{N}\int_0^\infty d\omega
I(\omega)
\left[1-\left(\frac{\omega}{\omega_0}\right)^2 \right]^{-2}
\nonumber \\
&\times&
\Bigg\{\left[1+\left(\frac{\omega}{\omega_0}\right)^2\right]
\left[1-\cos(\omega_0 t)  \cos(\omega t) \right]
\nonumber \\
&-&
2\frac{\omega}{\omega_0}\sin(\omega t)\sin(\omega_0 t)\Bigg\},
\label{Eq:PhaseAve}
\end{eqnarray}
where $N$ is the normalization factor. 
We average
also over the frequency, and thus the normalizing factor,
in the case we consider, becomes
\begin{equation}
N=\omega_c\frac{2}{\pi} \int_0^{\infty}d\omega~\frac{\omega_c^2}{\omega_c^2+\omega^2}=\omega_c^2.
\end{equation}

We obtain the result for the heating function dynamics of a damped
oscillator by plugging in the spectral distribution [Eq.~(\ref{Eq:IHt})],
into the phase averaged result, Eq.~(\ref{Eq:PhaseAve}).
In the regime $r<1$ the non-Lindblad-type dynamics dominates.
The environment is detuned from the system frequency, the most
intense part of the environment appears for small
frequencies $\omega\ll\omega_0$ whereas the on-resonant
part corresponds to the low intensity part of the spectral density. We set the parameters
as follows: $r=0.1,~g=0.045,~k_{\rm B}/\omega_0=80$.

The results from single frequency drives
[Figs.~\ref{Fig:Freqs} (a)-(c)]
suggest that the low frequency part of the spectrum
(part I in Fig.~\ref{Fig:Spectrum1}) induces
oscillations of the heating function.
In contrast, the result for $\omega=\omega_0$
[Fig.~\ref{Fig:Freqs} (d)] indicates that
the near-resonant part of the environment spectrum
(part II in Fig.~\ref{Fig:Spectrum1}),
produces a Markovian-type heating. 

Indeed, when we calculate the result by integrating over
the low frequency part, $0\leqslant\omega/\omega_0\leqslant0.2$,
the heating function displays damped oscillations, see Fig.~\ref{Fig:Parts} (a).
When the result is calculated by integrating over the on-resonant
part, $0.8\leqslant\omega/\omega_0\leqslant1.2$, the heating
function displays linear dynamics, see Fig.~\ref{Fig:Parts} (b).
When we add these two results
obtained by integrating over the limited frequency range,
the heating function displays the dominant characteristics
of the exact result and the match between the two results
is already quite close, see Fig.~\ref{Fig:Parts} (c).
If we then add to the integration the gap between
the two frequency regimes presented above, the match
between the quantum simulator and the exact results is excellent
and it becomes difficult to see any difference
between the two, 
see Fig.~\ref{Fig:Parts} (d).

These results illustrate how we can identify the origin
of the dominating features of the heating function
dynamics in the various parts of the environment spectrum.
It is also worth noting that the simulator result in 
Fig.~\ref{Fig:Parts} (d) includes the non-adiabatic effects,
which are not contained in the simple result of Fig.~\ref{Fig:Parts} (c).
Obviously, the inclusion of the non-adiabatic effects
enhances the match giving excellent agreement between
the quantum simulator and the exact results.
Moreover, according to the simple displacement
scheme for single frequency drives
(presented in the previous section),
the damping of the heating function oscillations 
arises in a subtle way as an indirect
consequence of several low frequencies
involved. All the low frequencies involved affect
the simulator dynamics in a similar way when
the driving field is switched on. However, the effect
of the second displacement, when the field is switched
off, depends on the frequency and 
it consequently damps the oscillations.

We have also checked the agreement between the exact and quantum simulator
results for the typical quadratic non-Markovian regime ($r\sim10$) and
for a Markovian regime ($r>20,~t\gg1/\omega_c$). For a fundamental theoretical study
of various types of dynamics see Ref.~\cite{Maniscalco04a}.

It is also interesting to study the case where the phase
of the driving field is kept fixed. This illuminates an 
aspect about the effect of the random phase of the environmental noise
being simulated.
The result for the fixed-phase, $\varphi=0$, is displayed in 
Fig.~\ref{Fig:Parts} (d), dotted line.
The random phase does not seem to play
a role for Markovian heating since the slope of the linear
increase in $\langle n \rangle$ for large times is the same
for the fixed phase and for the phase averaged results.
Instead, the phases of the low frequency fields affect
the short time oscillatory behavior of the
heating function by influencing the
amplitude of the oscillation.
Without averaging over the phase,
and setting $\varphi=0$, gives larger oscillations
as can be seen in Fig.~\ref{Fig:Parts} (d).
When studying the Markovian case ($r>20$), we noticed,
that the fixed phase and random phase
results very closely agree. This indicates again 
that the random phase plays a significant role only
in the non-Markovian dynamics.

\subsubsection{Density matrix}
So far we have demonstrated that the averaged driven oscillator results can mimic correctly the heating function dynamics of the open system. It remains to be shown that 
actually the simulated density matrix matches the density matrix of the non-Markovian
damped harmonic oscillator, given by Eq.~(\ref{Eq:RhokkAna}).

The single frequency drive (see Eq.~(\ref{Eq:H}) for the Hamiltonian) creates a coherent state. Following Ref.~\cite{Gardiner96a} and using the complex field representation
of the driving force, it can be shown
that the amplitude of the coherent state, for driving field frequency $\omega$ and
phase $\varphi$, is given by 
\begin{eqnarray}
\beta (t, \omega, \varphi) &=& -\kappa\left\{\frac{e^{-i\varphi}}{\frac{\omega}{\omega_0} + 1}
\left[ e^{i\omega_0t} - e^{-i\omega t}\right] 
\right. \nonumber \\
&+& \left.  \frac{e^{i\varphi}}{\frac{\omega}
{\omega_0} - 1} \left[ e^{i\omega t} - e^{i\omega_0 t}\right]\right\},
\label{Eq:Beta}
\end{eqnarray}
where the dimensionless coupling constant $\kappa$ is
\begin{equation}
\kappa=\frac{A}{\sqrt{2m\hbar\omega_0^3}}.
\end{equation}
The matrix elements
in the Fock-state basis are given in the usual way
\begin{equation}
\rho_{\rm k,l}(t, \omega, \varphi) =  e^{-|\beta|^2} \frac{\beta^k\beta^{\ast l}}{\sqrt{k! l!}}
= e^{-|\beta|^2} \frac{\beta^{k-l}|\beta|^{2l}}{\sqrt{k! l!}}.
\label{Eq:Rhokl}
\end{equation}
We recall that the analytic result for the damped harmonic oscillator
is valid for weak couplings.
To compare the results, we expand the exponential in Eq.~(\ref{Eq:Rhokl}) and 
keep the terms to second order
in the dimensionless coupling constant $\kappa$.
For the sake of simplicity we consider here the case of an initial state
$|n=0\rangle$ but the results can be generalized in a straightforward way
to other initial Fock states.
Moreover, since we are interested in the non-Markovian dynamics,
we look for the dynamics for times much shorter than the thermalization time.
With these conditions the diagonal elements of the density matrix,
up to second order in $\kappa$, are
\begin{eqnarray}
\rho_{0,0} (t, \omega, \varphi) &=& 1 - |\beta(\omega, \varphi, t)|^2 \nonumber \\
\rho_{1,1} (t, \omega, \varphi) &=& |\beta(\omega, \varphi, t)|^2 \nonumber \\
\rho_{k,k} (t, \omega, \varphi) &=& 0,~k\geqslant2,
\end{eqnarray}
while the off-diagonal elements are
\begin{eqnarray}
\rho_{1,0} (t, \omega, \varphi) &=& \rho_{0,1}^{\ast} = \beta(\omega, \varphi, t) \nonumber \\
\rho_{2,0} (t, \omega, \varphi) &=& \rho_{0,2}^{\ast} = \beta^2(\omega, \varphi, t).
\label{Eq:Rho20}
\end{eqnarray}
Similarly to the heating function calculation, the final result is obtained by averaging over the phase and frequency of the driving field.
Remembering that $I(\omega)/N$ is the normalized weight in the averaging
and $\langle n \rangle (t, \omega, \varphi)= |\beta(t, \omega, \varphi)|^2$,
with $\langle n \rangle (t, \omega, \varphi)$ 
given by Eq.~(\ref{Eq:WithPhi}), we obtain
\begin{eqnarray}
\rho_{0,0} (t) &=&  1- \frac{1}{N}\int_0^\infty d\omega 
\frac{1}{2\pi}\int_0^{2\pi}d\varphi
I(\omega)|\beta(t, \omega,\varphi)|^2
\nonumber \\
 &=&  1- \langle n(t) \rangle \nonumber \\
\rho_{1,1} (t) &=&  \frac{1}{N}\int_0^\infty d\omega 
\frac{1}{2\pi}\int_0^{2\pi}d\varphi
I(\omega)|\beta(t, \omega,\varphi)|^2 
\nonumber \\
&=& \langle n(t) \rangle.
\label{Eq:SRho}
\end{eqnarray}
Since we know that the heating function of the simulator matches
the one of the damped harmonic oscillator we can conclude that
the diagonal elements given by Eq.~(\ref{Eq:SRho})
coincide with Eq.~(\ref{Eq:AnaRho}). In order to prove that
the state of the simulator is the time dependent thermal state of
Eq.~(\ref{Eq:RhokkAna}), now we have to prove that the out of diagonal elements
are equal to zero. 

The phase dependence of
$\rho_{1,0} (t, \omega, \varphi)$
appears in the exponentials $\exp(\pm i\varphi)$, see Eq.~(\ref{Eq:Beta}).
Averaging over the phase therefore gives  
\begin{equation}
\rho_{1,0} (t) \propto \int_0^{2\pi} e^{\pm i\varphi}=0.
\end{equation}
It is worth reminding that Eq.~(\ref{Eq:MQbm}) describes the non-Markovian dynamics of the damped harmonic oscillator in the secular approximation. The same approximation in the simulator case amounts at averaging to zero the rapidly oscillating terms appearing in $\rho_{2,0}(t)$, see Appendix B.  Therefore, in the secular approximation,
$\rho_{2,0}(t)\simeq0$. Summarizing, we have proved that, in the secular approximation, the density matrix of the simulator coincides with the density matrix of the damped harmonic oscillator.

\begin{figure}[tb]
\centering
\includegraphics[scale=0.4]{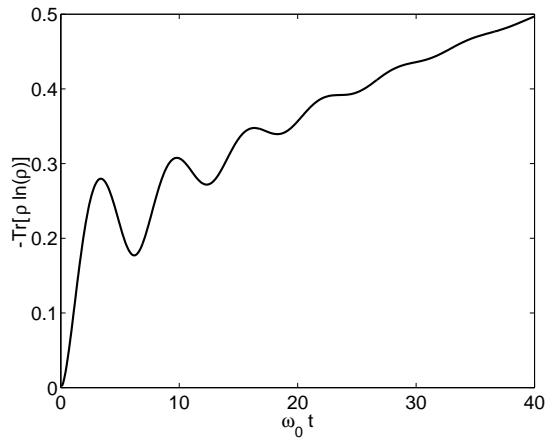}
\caption{\label{Fig:Entropy} 
The entropy production of the simulated damped oscillator in an engineered reservoir
for $r=0.1$. The oscillations of entropy directly reflect the inherent oscillatory
behavior of the populations and of the heating function for the short time. 
}
\end{figure}

It is worth noting that, since the density matrix of the damped harmonic oscillator and the one of the simulator approach coincide, the averaged driven closed system mimics
correctly also the entropy production typical for open quantum systems. In other words, 
while the closed driven oscillator remains
always in a pure state (no entropy production)  
the averaged state, i.e.~the state of the simulator, is a mixed state since Tr$\rho^2 < 1$ for time $t>0$ and allows the entropy production.
To illustrate this point further, we display in Fig.~\ref{Fig:Entropy}, by using the simulator result given by Eq.~(\ref{Eq:SRho}), the von Neumann entropy of the system
$S = -Tr(\rho \ln \rho)$. The oscillatory behavior of the entropy is a direct consequence of the oscillatory behavior of the heating function. For a more detailed description for the various types of heating function dynamics for engineered reservoirs, see Ref.~\cite{Maniscalco04a}. 

%%%%%%%%%%%% FIG
%%%%%%%%%%%% FIG

\section{Implementing the simulator with a driven trapped ion}\label{Sec:IonImp}

Single trapped ions are a well known example of a realization
of the quantum harmonic oscillator~\cite{Wineland2}. Recently these systems
have also been exploited to demonstrate the creation of artificial
engineered reservoirs for open quantum systems~\cite{ReservoirEng,WinelandPRA}.
The ability to cool the ion to its ground vibrational state
and the long coherence time
make this system ideal for the implementation of the
quantum simulator described in this paper.

Important factors for the implementation of
the quantum simulator presented here 
are the number and frequency range of the driving fields
to be used in the experiment.
Suppose now that we want to simulate the effect of non-Markovian off-resonant engineered reservoir to reveal
the characteristic quantum mechanical oscillatory behavior of the oscillator
heating function. We set the parameters
to correspond to Fig.~\ref{Fig:Parts},
$r=0.1,~\omega_c/2\pi=1.1$ MHz and $k_{\rm B}T/\omega_0=80$.

The experimental procedure is the following. We
keep the step size of the frequency sampling fixed
and apply a conveniently small number of pulses. We treat the phase
of the field as a random variable.
A typical oscillator frequency of the trapped ion in current
experiments is $\omega_0/2\pi=11$ MHz~\cite{WinelandPRA}.
The frequency range of the periodic drive  
is, e.g.,~$0-12.3$ MHz and the separation
between the frequencies of the sets of drives
 is $55$ kHz.
 In this example the total number of different applied frequencies
 is $225$.
For each frequency, the sets of different duration
have to be applied in order to obtain the
time evolution.
The result for the heating function
is obtained by converting the integral of Eq.~(\ref{Eq:GenAve})
to a sum over the sets of applied electric pulses.
The result of the summation has excellent
agreement with the exact and theoretical
quantum simulator results of Fig.~\ref{Fig:Parts} (d),
and to avoid repetition we have not plotted
in this figure the third overlapping curve.
Instead it is worth mentioning that the x-axis of the figure,
for the chosen experimental parameters, corresponds to
 $\mu$s time scale.

The system-reservoir coupling constant is directly proportional
to the intensity of the applied field and the dimensioless
coupling constant appearing in the analytical solution
is directly given by
\begin{equation}
g^2 = \kappa^2=\frac{e^2 E^2}{2m\hbar\omega_0^3},
\end{equation}
where $E$ is the amplitude of the applied field
and $e$ the charge of the ion.
Actually, one can choose 
the intensity for each set of drives $\omega$ in the most 
convenient way for the experimental implementation.
This is because the used intensities
can  be rescaled when calculating the average
over the frequencies for the heating function. In other words, the intensity dependence
$|\alpha|^2\propto E^2$ in Eq.~(\ref{Eq:PhaseAve})
stands outside the integration and can be taken outside of 
the corresponding summation of the experimental
results. 
It is also worth noting that the ambient reservoir
does not disturb the experiment since its effect occurs
on few orders of magnitude larger time scale~\cite{Wineland2}.
The vibrational
quantum number of the ion can be measured in the
standard way~\cite{Wineland2}. 

The displacement scheme presented in 
Sec.~\ref{Sec:Drive} has
straightforward implications for interpreting
the ion simulator results.
A sudden switch on and off of the electric field displaces
the centre of the ion trap due to interaction
between the charge of the ion and the electric field.
The central point is that in the adiabatic regime,
the center of the trap moves slowly compared to the
oscillator frequency, and the ion follows adiabatically
the motion of the trap center. This effect plays
a key role in the damping of the heating function
in the experiment.

\section{Discussion and conclusions}\label{Sec:DisCon}

In the past, the average over the stochastic fields has been used
to describe the effects of ambient reservoirs to the oscillator dynamics,
e.g.~the heating of the ground state of trapped 
ions~\cite{Lamoreaux,James,Budini}.
Here, we use similar displacement operator based formalism
to investigate a new aspect to the topic. We
ask the opposite question:
how can controlled drives and their average be used 
to model the effects of a reservoir?  In fact, we demonstrate how to use controlled
drives to mimic the effects of both the ambient and
the artificial engineered reservoirs.

In addition of the developed quantum simulator,
the used approach brings to light some
interesting features
of the open system dynamics. 
We have been dealing with
a characteristic quantum mechanical non-Lindblad-type
master equation~\cite{Maniscalco04a,Maniscalco04c}. 
Thus, it is somewhat surprising that the oscillations of the
heating function, a typical quantum mechanical
dynamical feature, have a semiclassical
interpretation, when the simulator is implemented
with a trapped ion. Namely, the source of the oscillations,
according to our interpretation, can be seen
in the \lq\lq displacement--coherent state dynamics--displacement\rq\rq~
scheme described in Sec.~\ref{Sec:Drive}. In contrast, usually the oscillatory
behavior of the heating function and non-Lindblad-type--dynamics
are connected to the appearance
of virtual processes~\cite{Maniscalco04a}. 

The results also
illuminate the role of the random phase of the environment fields.
In the case of oscillatory non-Markovian dynamics, the role of the
random phase is to affect the amplitude of the short time oscillations.
The oscillation is initiated by the low frequency fields
and its amplitude is 
given by an average over the effects of the random phase of the same
frequencies. The oscillations are then damped because of the
various frequencies involved.
For strictly Markovian case, the phase plays
hardly any role. 
 
Our results also shed some light on the problem of the existence of a
continuos measurement scheme interpretation of the quantum
trajectories for non-Lindbad-type dynamics.
In the Lindblad case there is a direct correspondence between the 
continuos measurement scheme of the environment
and the quantum trajectories generated by the computer 
simulations~\cite{MeasurementScheme}.
For the non-Lindblad case
a completely satisfactory  connection is still lacking 
despite of some attempts~\cite{Gambetta1,Breuer2},
and this important problem
remains open. Thus, it is interesting to note that,
in order to observe the oscillations of the heating function
with the quantum simulator described here,
the act of switching off the field
(second displacement) has a crucial role.
No measurements are allowed 
while the driving field is on. 
This points to the direction that it is probable
that a satisfactory measurement scheme interpretation
for non-Lindblad type dynamics does not exist,
at least not in the framework of conventional quantum theory.
It might be interesting to study in the future what the  
measurement
scheme interpretations based
on alternative formulations of quantum mechanics~\cite{Gambetta1}
would say in the context of the quantum simulator proposed here.

It is important to note that most of the existing analytical methods
to study open system dynamics rely on various
approximations, e.g.~weak coupling
between the system and the environment~\cite{Breuer}. It
would be extremely stimulating for the research
on open quantum systems to possess a general
quantum simulator which could be used
in the regimes where most of the approximations
used in the analytical calculations fail. We have taken
an initial step towards this direction and a more
general quantum simulator for open systems
will be a challenging task for our future studies.

In conclusion, we have shown how to use a controlled
driven harmonic oscillator to simulate the behavior of a damped oscillator. In other
words, our results demonstrate the  possibility of studying one of the 
paradigmatic open quantum systems models by means of a closed quantum
system. Moreover, we have discussed in detail the implementation of
the quantum simulator with single trapped ions.
In addition, the simulator approach illuminates several interesting aspects
of the non-Markovian dynamics of a damped oscillator,
most notably identifying the role that the various parts
of the non-Markovian spectrum 
play in the system dynamics.

\acknowledgments

The authors thank D. Wineland for stimulating exchange
of ideas 
and K.-A. Suominen
for illuminating discussions.
This work has been supported by the Academy of Finland 
(projects 207614, 206108, 108699), the Magnus
Ehrnrooth Foundation, and the European Union's Transfer of Knowledge
project CAMEL (Grant No. MTKD-CT-2004-014427).

\appendix

\section{}

In this Appendix A we show that for an initial Fock state
$|n=0\rangle$ the state of the system as a function of time is given by Eq.~(\ref{Eq:RhokkAna}).

The quantum characteristic function (QCF) of the system at time $t$ can
be written \cite{Intravaia03c} (for a general definition of QCF 
see, e.g.,~Ref.~\cite{Barnett})
\begin{equation}
\chi_t (\xi)=e^{- \Delta_{\Gamma}(t)|\xi|^2} \chi_0 \left(
e^{- \Gamma (t)/2} \xi   \right),
\label{chit}
\end{equation}
where $\chi_0$ is the QCF of the initial state of the system, and
the quantities $\Delta_{\Gamma}(t)$ and $\Gamma(t)$
are defined in terms of the diffusion and dissipation
coefficients $\Delta(t)$ and $\gamma(t)$ respectively [see
master equation (\ref{Eq:MQbm})] as follows
\begin{eqnarray}
\Gamma(t)&=& 2\int_0^t \gamma(t_1)\:dt_1, \label{Gamma} \\
\Delta_{\Gamma}(t) &=& e^{-\Gamma(t)}\int_0^t
e^{\Gamma(t_1)}\Delta(t_1)dt_1 \label{DeltaGamma}.
\end{eqnarray}
Equation~(\ref{chit}) shows that the QCF is the product of an
exponential factor and a transformed
initial QCF. 

For a Fock state $|n=0\rangle$ the initial QCF is
\cite{Barnett}
\begin{equation}
\chi_0(\xi') = e^{-|\xi'|^2/2}.
\label{Eq:ZeroT}
\end{equation}
By plugging Eq.~(\ref{Eq:ZeroT}) into Eq.~(\ref{chit}) with $\xi'=e^{- \Gamma (t)/2} \xi$,
the QCF at time $t$
can be written
\begin{equation}
\chi_t (\xi)=\exp\left\{- \left[ \Delta_{\Gamma}(t) + \frac{1}{2}e^{-\Gamma (t)} \right] |\xi|^2  \right\}.
\label{chit2}
\end{equation} 

The heating function in turn, for an intial ground state of the system, can be 
written \cite{Maniscalco04a,Maniscalco04c}
\begin{equation}
\langle n (t) \rangle = \frac{1}{2} \left( e^{-\Gamma(t)}-1\right)+
\Delta_{\Gamma}(t),
\end{equation}
which gives 
\begin{equation}
\langle n (t) \rangle +\frac{1}{2}= \Delta_{\Gamma}(t) + \frac{1}{2} e^{-\Gamma(t)}.
\label{Eq:Ntt}
\end{equation}
 The right hand side of Eq.~(\ref{Eq:Ntt}) appears in the expression 
for the QCF in Eq.~(\ref{chit2}) which can consequently be written
\begin{equation}
\chi_t (\xi)=\exp \left\{ - \left[ \langle n (t) \rangle +\frac{1}{2} \right] |\xi|^2 \right\}.
\label{chit3}
\end{equation}
It is also worth mentioning that for high temperature
reservoir ($k_BT/\omega_0\gg 1$) the 
dissipation coefficient $\gamma(t)\ll \Delta(t)$ and the heating function
dynamics can be approximated with Eq.~(\ref{Eq:Exact})~\cite{Maniscalco04a}.

The general expression for a QCF of a thermal state is
\cite{Barnett}
\begin{equation}
\chi(\xi)=\exp \left\{ - \left[  \bar{n} +\frac{1}{2} \right] |\xi|^2 \right\},
\label{Eq:Thermal}
\end{equation}
where $ \bar{n}$ is the mean excitation number of the thermal state.
Comparison between the QCF of the system, Eq.~(\ref{chit3}), and
the QCF of the thermal state, Eq.~(\ref{Eq:Thermal}), shows that the
state of the system at each time, for the initial state we consider,
is a thermal state with changing \lq\lq effective\rq\rq~temperature given
by the heating function  $\langle n (t) \rangle$. Equation (\ref{Eq:RhokkAna}) follows
then directly from this observation.

We note also that 
the time dependent 
\lq\lq effective\rq\rq~temperature
coincides with the temperature of the environment
only in the long time limit when the system has reached a thermal
equilibrium with its environment.
It is also worth mentioning that the QCF of the thermal state
with ${\bar n}=0$, Eq.~(\ref{Eq:Thermal}), matches the QCF of the
Fock state $|n=0\rangle$, Eq.~(\ref{Eq:ZeroT}), since
the Fock-state $|n=0\rangle$ coinsides with the $T=0$
thermal state.

\section{}

In this Appendix B we show that in the secular approximation 
$\rho_{2,0}(t)\simeq 0$.  From Eqs. (21) and (25) one derives
\begin{eqnarray}
\rho_{2,0}(t) \propto \int_0^{\infty} d \omega \int_0^{2 \pi} d \varphi
I(\omega) [\beta_1(t,\omega,\varphi)+
\beta_2(t,\omega,\varphi)]^2, \nonumber \\ 
&&
\end{eqnarray}
with
\begin{eqnarray}
\beta_1(t,\omega,\varphi) &\propto& \frac{e^{-i
\varphi}}{\omega+\omega_0}\left( e^{i\omega_0 t}- e^{-i\omega t}
\right)
\\
\beta_2(t,\omega,\varphi) &\propto& \frac{e^{i
\varphi}}{\omega-\omega_0}\left( e^{i \omega t} - e^{i\omega_0 t}
\right).
\end{eqnarray}
We note that
\begin{eqnarray}
&& \int_0^{2\pi} d\varphi\left[ \beta_1^2(t,\omega,\varphi)+
\beta_2^2(t,\omega,\varphi) \right] \nonumber \\  
 &=&
 \int_0^{\pi}
d\varphi ( c_1 e^{-2i \varphi}+ c_2 e^{2 i \varphi}) =0,
\end{eqnarray}
where $c_{1,2}$ are functions which do not depend on $\varphi$.
Therefore
\begin{eqnarray}
&&\rho_{2,0}(t)\propto \int_0^{\infty} d \omega \int_0^{2 \pi} d \varphi
I(\omega) \beta_1(t,\omega,\varphi) \beta_2(t,\omega,\varphi)
\nonumber \\
&&\propto \int_0^{\infty} d \omega I(\omega) \frac{e^{i \omega_0
t}}{\omega^2-\omega_0^2}
\sin[(\omega-\omega_0)t]\sin[(\omega+\omega_0)t]\nonumber \\
&&\propto e^{i \omega_0 t} \left[e^{-2 \omega_c t} +
\left(\frac{2i \omega_c}{\omega_0}-1 \right)\cos(2 \omega_0 t)+
\frac{\omega_c}{\omega_0} \sin (2 \omega_0 t) \right].\nonumber \\
&&
\end{eqnarray}
In the secular approximation, used for the study of the damped
harmonic oscillator, the rapidly oscillating terms appearing in
the previous equation average out to zero, as one can easily see
plotting the real and imaginary part of $\rho_{2,0}(t)$.

{}


\begin{thebibliography}{99}

\bibitem{Nielsen}
M. A. Nielsen and I. L. Chuang, {\it Quantum Computation and
Quantum Information} (Cambridge University Press, Cambridge, 2000).

\bibitem{Stenholm}
S. Stenholm and K.-A. Suominen,
{\it Quantum Approach to Informatics}
(Wiley, New York, 2005).

\bibitem{EuDocu}
P. Zoller {\it et al.}, Eur. Phys. J. D {\bf 36}, 203 (2005).

\bibitem{Sorensen}
A. S{\o}rensen and K. M{\o}lmer,
Phys. Rev. Lett. {\bf 83}, 2274 (1999).

\bibitem{Wineland1}
D. Leibfried {\it et al.}, Phys. Rev. Lett. {\bf 89}, 247901 (2002).

\bibitem{Zoller1}
H. P. B\"{u}chler, M. Hermele, S. D. Huber, M. P. A. Fisher,
and P. Zoller, Phys. Rev. Lett. {\bf 95}, 040402 (2005).

\bibitem{Milburn}
P. M. Alsing, J. P. Dowling, and G. J. Milburn,
Phys. Rev. Lett. {\bf 94}, 220401 (2005).

\bibitem{Stoof}
M. Snoek, M. Haque, S. Vandoren, and H. T. C. Stoof,
Phys. Rev. Lett. {\bf 95}, 250401 (2005).

\bibitem{Wineland2}
D. Leibfried, R. Blatt, C. Monroe, and D. Wineland,
Rev. Mod. Phys. {\bf 75}, 281 (2003).

\bibitem{Breuer}
H.-P. Breuer and F. Petruccione, {\it The Theory of Open
Quantum systems}  (Oxford University Press,  Oxford, 2002).

\bibitem{Qo}
L. Mandel and E. Wolf, {\it Optical Coherence and Quantum Optics}
(Cambridge University Press, Cambridge, 1995)

\bibitem{Np}
I. Joichi, Sh. Matsumoto, and M. Yoshimura,
Phys. Rev. A {\bf 57},  798 (1998).

\bibitem{Ch}
P. H\"anggi, P. Talkner, and M. Borkovec,
Rev. Mod. Phys. {\bf 62}, 251 (1990).

\bibitem{Lamoreaux}
S. K. Lamoreaux,
Phys. Rev. A {\bf 56}, 4970 (1997).

\bibitem{James}
D. F. V. James, 
Phys. Rev. Lett. {\bf 81}, 317 (1998).

\bibitem{Budini}
A. A. Budini, Phys. Rev. A {\bf 64}, 052110 (2001).

\bibitem{Zoller2}
J. F. Poyatos, J. I. Cirac, and P. Zoller, Phys. Rev. Lett.
{\bf 77}, 4728 (1996).

\bibitem{ReservoirEng}
C. J. Myatt, B. E. King, Q. A. Turchette, C. A. Sackett,
D. Kielpinski, W. M. Itano, and D. J. Wineland,
Nature {\bf 403}, 269 (2000). 

\bibitem{WinelandPRA}
Q. A. Turchette,  C. J. Myatt, B. E. King, C. A. Sackett, D.
Kielpinski, W. M. Itano, C. Monroe, and D. J. Wineland,
Phys. Rev. A {\bf 62}, 053807 (2000).

\bibitem{Zoller3}
P. Rabl, A. Shnirman, and P. Zoller, Phys. Rev. B {\bf 70}, 205304 (2004).

\bibitem{MeasurementScheme}
M. B. Plenio and P. L. Knight,
Rev. Mod. Phys. {\bf 70}, 101 (1998),
and references therein.

\bibitem{Gardiner96a}
C. W. Gardiner and P. Zoller,
{\it Quantum Noise: A Handbook of Markovian
and non-Markovian Quantum Stochastic Methods
with Applications to Quantum Optics} (Springer-Verlag, Berlin, 1999).

\bibitem{Maniscalco04a}
S. Maniscalco, J. Piilo, F. Intravaia, F. Petruccione,
and A. Messina, Phys. Rev. A {\bf 70}, 032113 (2004).

\bibitem{Intravaia03a}
F. Intravaia, S. Maniscalco, and A. Messina,
Eur. Phys. J. D {\bf 32} 97 (2003).

\bibitem{Caldeira83a}
A. O. Caldeira and A. J. Leggett, Physica A {\bf 121}, 587 (1983).

\bibitem{Weiss}
U. Weiss, {\it Quantum Dissipative Systems}
(World Scientific Publishing, Singapore, 1999).

\bibitem{Maniscalco04c}
S. Maniscalco, J. Piilo, F. Intravaia, F. Petruccione, and A. Messina, 
Phys. Rev. A {\bf 69}, 052101 (2004). 

\bibitem{OscNote}
The physical explanation
of the oscillations of the heating function in the non-Markovian regime
 [Fig.~3 (d)] is the virtual exchange of energy
between the system and the environment. These have been
studied in more detail in Ref.~\cite{Maniscalco04a}.

\bibitem{Intravaia03b}
F. Intravaia, S. Maniscalco, J. Piilo, and A. Messina,
Phys. Lett. A {\bf 308}, 6 (2003).


\bibitem{Gambetta1} 
J. Gambetta and H. M. Wiseman, Phys. Rev. A {\bf 68}, 062104 (2003).

\bibitem{Breuer2}
H.-P. Breuer, Phys. Rev. A {\bf 70}, 012106 (2004).

\bibitem{Intravaia03c}
F. Intravaia, S. Maniscalco, and A. Messina,
Phys. Rev. A {\bf 67}, 042108 (2003).

\bibitem{Barnett}
S. M. Barnett and P. M. Radmore, 
{\it Methods in Theoretical Quantum Optics} (Oxford University Press, Oxford, 1997).

\end{thebibliography}
\end{document}